
\input epsf.tex

\font\rmu=cmr10 scaled\magstephalf
\font\bfu=cmbx10 scaled\magstephalf

\font\it=cmti10 scaled \magstephalf
\font\bf=cmbx10 scaled\magstephalf
\rmu

\font\rmus=cmr8
\font\rmuss=cmr6
\font\mait=cmmi10 scaled\magstephalf
\font\maits=cmmi7 scaled\magstephalf
\font\maitss=cmmi7
\font\msyb=cmsy10 scaled\magstephalf
\font\msybs=cmsy8 scaled\magstephalf
\font\msybss=cmsy7
\font\bfus=cmbx7 scaled\magstephalf
\font\bfuss=cmbx7
\font\cmeq=cmex10 scaled\magstephalf

\textfont0=\rmu
\scriptfont0=\rmus
\scriptscriptfont0=\rmuss

\textfont1=\mait
\scriptfont1=\maits
\scriptscriptfont1=\maitss

\textfont2=\msyb
\scriptfont2=\msybs
\scriptscriptfont2=\msybss

\textfont3=\cmeq
\scriptfont3=\cmeq
\scriptscriptfont3=\cmeq

\newfam\bmufam  \textfont\bmufam=\bfu
      \scriptfont\bmufam=\bfus \scriptscriptfont\bmufam=\bfuss

\hsize=15.5cm
\vsize=21cm
\baselineskip=16pt   
\parskip=12pt plus  2pt minus 2pt

\def\d{\delta}
\def\e{\epsilon}

\def\semi{\bigcirc\kern-1em{s}\;}

\def\ni{\noindent}

\def\one{{\mathchoice {\rm 1\mskip-4mu l} {\rm 1\mskip-4mu l}
{\rm 1\mskip-4.5mu l} {\rm 1\mskip-5mu l}}}
\def\Q{{\mathchoice
{\setbox0=\hbox{$\displaystyle\rm Q$}\hbox{\raise 0.15\ht0\hbox to0pt
{\kern0.4\wd0\vrule height0.8\ht0\hss}\box0}}
{\setbox0=\hbox{$\textstyle\rm Q$}\hbox{\raise 0.15\ht0\hbox to0pt
{\kern0.4\wd0\vrule height0.8\ht0\hss}\box0}}
{\setbox0=\hbox{$\scriptstyle\rm Q$}\hbox{\raise 0.15\ht0\hbox to0pt
{\kern0.4\wd0\vrule height0.7\ht0\hss}\box0}}
{\setbox0=\hbox{$\scriptscriptstyle\rm Q$}\hbox{\raise 0.15\ht0\hbox to0pt
{\kern0.4\wd0\vrule height0.7\ht0\hss}\box0}}}}
\def\C{{\mathchoice
{\setbox0=\hbox{$\displaystyle\rm C$}\hbox{\hbox to0pt
{\kern0.4\wd0\vrule height0.9\ht0\hss}\box0}}
{\setbox0=\hbox{$\textstyle\rm C$}\hbox{\hbox to0pt
{\kern0.4\wd0\vrule height0.9\ht0\hss}\box0}}
{\setbox0=\hbox{$\scriptstyle\rm C$}\hbox{\hbox to0pt
{\kern0.4\wd0\vrule height0.9\ht0\hss}\box0}}
{\setbox0=\hbox{$\scriptscriptstyle\rm C$}\hbox{\hbox to0pt
{\kern0.4\wd0\vrule height0.9\ht0\hss}\box0}}}}

\font\fivesans=cmss10 at 4.61pt
\font\sevensans=cmss10 at 6.81pt
\font\tensans=cmss10
\newfam\sansfam
\textfont\sansfam=\tensans\scriptfont\sansfam=\sevensans\scriptscriptfont
\sansfam=\fivesans
\def\sans{\fam\sansfam\tensans}
\def\Z{{\mathchoice
{\hbox{$\sans\textstyle Z\kern-0.4em Z$}}
{\hbox{$\sans\textstyle Z\kern-0.4em Z$}}
{\hbox{$\sans\scriptstyle Z\kern-0.3em Z$}}
{\hbox{$\sans\scriptscriptstyle Z\kern-0.2em Z$}}}}

\newcount\foot
\foot=1
\def\note#1{\footnote{${}^{\number\foot}$}{\ftn #1}\advance\foot by 1}

\def\frac#1#2{{#1\over #2}}
\def\text#1{\quad{\hbox{#1}}\quad}

\font\ch=cmbx12 scaled\magstephalf
\font\ftn=cmr8 scaled\magstephalf

\font\it=cmti10 scaled\magstephalf
\font\bf=cmbx10 scaled\magstephalf
\font\titch=cmbx12 scaled\magstep2
\font\titname=cmr10 scaled\magstep2
\font\titit=cmti10 scaled\magstep1
\font\titbf=cmbx10 scaled\magstep2

\nopagenumbers

\line{\hfil December 31, 1996}
\vskip1.5cm
\centerline{\titch FURTHER RESULTS ON GEOMETRIC OPERATORS}
\vskip.5cm
\centerline{\titch IN QUANTUM GRAVITY}
\vskip1.7cm
\centerline{\titname R. Loll}
\vskip.5cm
\centerline{\titit Max-Planck-Institut f\"ur Gravitationsphysik}
\vskip.2cm
\centerline{\titit Schlaatzweg 1}
\vskip.2cm
\centerline{\titit D-14473 Potsdam, Germany}
\vskip.3cm
\centerline{and}
\vskip.3cm
\centerline{\titit Erwin Schr\"odinger Institut}
\vskip.2cm
\centerline{\titit Boltzmanngasse 9}
\vskip.2cm
\centerline{\titit A-1090 Wien, Austria}

\vskip1.5cm
\centerline{\titbf Abstract}
\vskip0.2cm

We investigate some properties of geometric operators 
in canonical quantum gravity in the connection approach \`a la Ashtekar,
which are associated with volume, area and length of spatial regions. 
We motivate the construction of analogous discretized lattice 
quantities, compute various quantum commutators of the type [area,volume],
[area,length] and [volume,length], and find they are generally 
non-vanishing.

Although our calculations are performed mostly within a lattice-regularized 
approach, some are -- for special, fixed spin-network 
configurations -- identical with corresponding continuum computations.
Comparison with the structure of the discretized theory leads us to 
conclude that anomalous commutators may be a general
feature of operators constructed along similar lines within a continuum loop
representation of quantum general relativity. -- 
The validity of the lattice approach remains unaffected.

\vfill\eject
\footline={\hss\tenrm\folio\hss}
\pageno=1

\line{\ch 1 Introduction\hfil}

The introduction of geometric quantum operators, like those measuring 
areas and volumes, has proven fruitful in the study of non-perturbative
canonical gravity in 3+1 dimensions. 
We are referring here to attempts of defining a theory of canonical
gravity in a so-called loop representation, where the basic variables
are one-dimensional holonomies along curves in spatial three-slices 
$\Sigma$ [1].
The present paper deals with the case where the holonomy variables are
obtained by integrating a real $su(2)$-valued connection form $A$ 
on $\Sigma$. $A$ is part of a canonical Yang-Mills-type variable pair 
$(A,E)$, which is a real version [2] of the well-known Ashtekar 
variables [3].

In the study of pure gravity without matter coupling, the geometric 
operators {\it per se} are primarily of interest at the kinematical level,
that is, before the quantum diffeomorphism constraints have been 
imposed on wave functions. They are not observables in that they do not 
commute with those constraints. 
Because of their obvious geometric interpretation, and relatively simple
form in the quantum theory, they have been applied in a variety of contexts.
A volume operator was first studied in [4], and part of its spectrum
analyzed in [5,6,7]. It has been used in the construction of the quantum 
Hamiltonian in the real connection approach, both on the lattice [8],
and in a continuum regularization [9].
The area operator was also investigated in [4], and more complete 
versions of its spectrum later
given in [10,11]. It has been applied in estimating the radiation of black 
holes [12] and making contact with semi-classical geometries, obtained from 
coarse-graining quantum states [13].
 
In the continuum quantum theory, wave functions at the kinematical level
are labelled by so-called spin-network states [14], which are particular
totally anti-symmetrized linear combinations of Wilson loops (gauge-invariant
quantities with respect to local SU(2)-rotations, obtained by taking 
traces of holonomies of closed curves). This is simply a variation on
the old theme of loop representation on the space of connections modulo
gauge. The advantage of spanning the Hilbert space $\cal H$ of 
square-integrable
functions on this space by the spin networks is that in this basis 
the geometric operators can be diagonalized easily, namely, on 
finite-dimensional subspaces of $\cal H$.

Things look somewhat similar when one tries to construct a lattice
regularization of connection gravity. The most obvious ansatz is to
proceed as in Hamiltonian lattice gauge theory, and use a cubic lattice
with discretized Hamiltonian link variables [15,5,6]. In particular, one 
may go to the gauge-invariant sector of Hilbert space, and employ a basis
of spin-network states. The only significant difference with the continuum is
that the configuration space underlying the Hilbert space is 
finite-dimensional (for a finite lattice), and the lattice links (edges)
themselves are a discrete approximation of space, instead of being
imbedded in a given manifold $\Sigma$.

In the construction of geometric operators in the continuum, one roughly
speaking proceeds as follows (see, for example, [10]): 
first, one smears out the bare operators 
which are polynomials in $\hat E$ and therefore contain multiple 
derivatives at a point. There is no unique way of doing this, and we cannot 
comment here on the virtue of the different procedures people have adopted. 
One then defines the regularized
operator $\hat {\cal O}_{\rm reg}$ (describing the $n$-volume of some finite 
spatial region), evaluates it on a quantum state $\psi_\gamma$, and
looks at the entire expression $\hat{\cal O}_{\rm reg}\psi_\gamma$ in the 
limit as the relevant regularization parameters are taken to zero. 
The resulting expression is usually finite (i.e. no further renormalization
is necessary). This is closely related to the kind of quantum representation
one is considering, where typical quantum excitations are taken to
be finite linear combinations of spin-network states associated with
finite imbedded graphs $\gamma$ in $\Sigma$. The smeared-out operators act
non-trivially only at points $x\in\Sigma$ which happen to be crossed
by an edge of the graph $\gamma$ underlying $\psi_\gamma$. 

For the volume operators, the
action reduces to a sum over intersection points of the graph
that happen to lie in the given spatial region whose volume is to be
measured. For the area operators, it reduces to a sum over intersection
points of edges of the graph with the relevant two-surface whose area
is to be determined. For finite and well-behaved graphs these sums are
{\it finite}. Moreover, the ``remainder" of the operator action at each 
of the finite number of contributing points is rather simple: it
corresponds to a finite rearrangement of how the incoming flux lines
of $\psi_\gamma$ can be contracted gauge-invariantly at $x$.  
(A similar construction can be performed for the length operator,
but leads to the counter-intuitive result that the length of most
curves is zero. This happens because a one-dimensional piece of curve 
-- whose length is to be measured -- generically does not have any 
intersections with the set of vertices of a graph $\gamma\in\Sigma$. 
One way of ``fixing'' this problem by introducing a spatial smearing for 
the length operator is described in [16].)
The spectra of all volume, area and length operators investigated up
to now in the continuum are discrete.

In the lattice theory, one may define discretized analogues of the
geometric operators (see, for example, [6]). Their structure is very similar to 
that of the ``finite remainder" of the continuum
operators described in the previous paragraph. Many calculations one performs on
the lattice can be considered as coming from a continuum calculation
on a quantum state where the underlying graph happens to be a cubic 
lattice. This will be explained in more detail in the main part of the paper.
Differences in interpretation do however arise, since the lattice
theory is only a finite-dimensional approximation to the real theory,
which will only be attained in some infinite-volume limit where 
the details of the discretization become unimportant.

The remainder of this paper is organized as follows. In the next 
section we set up the lattice description and define various ways of 
discretizing volume, area and length functions on the lattice. In 
Sec.3, we compute various commutators of the corresponding geometric lattice 
operators and give some explicit examples of spin-network 
configurations where the commutators do not vanish. 
We also discuss a selection rule for states with non-negative volume. 
In Sec.4 we investigate some implications of the former 
result and explain why the presence of anomalous commutators 
in the lattice theory is natural, and show that the lattice commutators 
obtain their expected form in the limit as the lattice spacing $a$ is taken 
to zero. Our calculations imply that anomalous commutators are
also present in the continuum theory. This is worrying, since there no 
continuum limit is usually deemed necessary. 
We argue that the origin of non-commutativity is
not a quantum effect, but lies in the choice of 
non-local basic variables in the continuum quantum theory.

\vskip2cm

\line{\ch 2 Defining geometric lattice operators\hfil}

We start with a brief summary of the basic ingredients of Hamiltonian 
lattice gauge theory \`a la Kogut and Susskind [17]. 
For computational simplicity, we take the lattice $\Lambda$ to be a cubic 
$N^3$-lattice with periodic boundary conditions. The basic quantum operators
associated with each lattice link $l$ are a group-valued $SU(2)$-link
holonomy $\hat V$ (represented by multiplication by $V$),
together with its inverse $\hat V^{-1}$, and a pair
of canonical momentum operators $\hat p^+_i$ and $\hat p^-_i$, where
$i$ is an adjoint index. The operator $\hat p^+_i(n,\hat a)$ is based
at the vertex $n$, and is associated with the link $l$ oriented in the
positive $\hat a$-direction. By contrast, $\hat p^-_i(n+\hat 1_{\hat
a},\hat a)$ is based at the vertex displaced by one lattice unit in
the $\hat a$-direction, and associated with the inverse link
$l^{-1}(\hat a)=l(-\hat a)$. In mathematical terms, the momenta 
$\hat p^+$ and $\hat p^-$ correspond to the left- and right-invariant
vector fields on the group manifold associated with a given link.
The wave functions are elements of
$\times_l L^2(SU(2),dg)$, with the product taken over all links, and
the canonical Haar measure $dg$ on each copy of the group $SU(2)$. 
The basic commutators are

$$
\eqalign{
&[\hat V_A{}^B(n,\hat a),\hat V_C{}^D(m,\hat b)]=0,\cr
&[\hat  p^+_i(n,\hat a),\hat V_A{}^C(m,\hat b)]=
-\frac{i}{2}\,\d_{nm}\d_{\hat a\hat b}\, \tau_{iA}{}^B\hat V_B{}^C
(n,\hat a),\cr
&[\hat  p^-_i(n,\hat a),\hat V_A{}^C(m,\hat b)]=
-\frac{i}{2}\,\d_{n,m+1}\d_{\hat a\hat b}\,\hat V_A{}^B
(n,\hat a) \tau_{iB}{}^C,\cr
&[\hat p^\pm_i(n,\hat a),\hat p^\pm_j(m,\hat b)]=
\pm i\, \d_{nm}\d_{\hat a\hat b}\, \e_{ijk}\, \hat p_k^\pm
(n,\hat a),\cr
&[\hat p^+_i(n,\hat a),\hat p^-_j(m,\hat b)]=0,}\eqno(2.1)
$$

\noindent where $\epsilon_{ijk}$ are the structure constants of $SU(2)$. 
The normalization for the $SU(2)$-generators $\tau_i$ is such that
$[\tau_i,\tau_j]=2\,\epsilon_{ijk}\tau_k$ and $A_a=A_a^i \tau_i/2$.

In order to relate discrete lattice expressions with their continuum
counterparts, one uses power series expansions in the so-called
lattice spacing $a$, which is an unphysical parameter with dimension
of length. For the basic classical lattice variables, these are

$$
\eqalign{
V_A{}^B(\hat b)&=1_A{}^B+a\,G\, A_{bA}{}^B+O(a^2),\cr
p_i^\pm (\hat b)&=a^2\,G^{-1}\, E^b_i +O(a^3).}\eqno(2.2)
$$

Note that Newton's constant $G$ appears since the dimensions of the
basic gravitational variables $A$ and $E$ differ from those of the
corresponding Yang-Mills phase space variables. We choose the 
components of the metric to be dimensionless, $[g_{ab}]=L^0$, which
leads to $[A_{grav}]=L^{-3}$ and $[E_{grav}]=L^0$, as opposed to
the usual $[A_{YM}]=L^{-1}$ and $[E_{YM}]=L^{-2}$ in gauge theory.

Using the expansions (2.2), one obtains unambiguous continuum 
limits of composite
classical lattice expressions by extracting the coefficient of the
lowest-order term in the $a$-expansion. The converse is not true:
there is no unique lattice discretization of a continuum expression,
since one may always add to the lattice version terms of higher
order in $a$ which do not contribute in the continuum limit. In
practice, one's choice of a lattice operator is usually motivated
by simplicity and considerations of symmetry, and will typically affect
the convergence behaviour of the (quantum) theory.

Following this prescription, let us now write down the lattice
equivalents of the volume, area and length functions.
In the continuum theory, these are simply spatial integrals of the square
root of the determinant of $g_{ab}$ (or the determinant of the induced
metric on the relevant 2- or 1-dimensional submanifold),

$$
\eqalign{
&{\cal V}^{\rm cont}({\cal R})=\int_{\cal R} d^3x \sqrt{g}=
\int_{\cal R} d^3x \sqrt{\frac{1}{3!}
\epsilon_{abc}\,\epsilon^{ijk} E^a_i E^b_j E^c_k},\cr
&{\cal A}^{\rm cont}({\cal S})=\int_{\cal S} d^2x \sqrt{{}^{(2)} g}=
\int_{\cal S} d^2x \sqrt{E^{3i} E^3_i},\cr
&{\cal L}^{\rm cont}({\cal C})=\int_{\cal C} dx \sqrt{{}^{(1)} g}=
\int_{\cal C} dx \sqrt{\frac{1}{\det E} (E^2_j E^{2j} E^3_k E^{3k}-
(E^{2j}E^3_j)^2\, ) },
}\eqno(2.3)
$$

\noindent where for simplicity we have chosen the surface ${\cal S}$
to be normal to the 3-direction everywhere and the curve $\cal C$ to lie
along the 1-direction. We are not concerned here with how the subspaces 
$\cal R$, $\cal S$, $\cal C$ of $\Sigma$ are defined (for example, they
could be determined through some matter distribution), since our
discussion in any case is restricted to the kinematical theory, where the
diffeomorphism symmetry has not yet been taken into account. 
As in [6], let us define the lattice function $D(n)$ as

$$
\eqalign{
D(n)&:=\epsilon_{abc}\,\epsilon^{ijk}p_i(n,\hat a)p_j(n,\hat b)
p_k(n,\hat c)\cr
&:=\frac{1}{8}\, \epsilon_{abc}\,\epsilon^{ijk}
(p^+_i(n,\hat a) +p^-_i(n,\hat a))
(p^+_j(n,\hat b) +p^-_j(n,\hat b))
(p^+_k(n,\hat c) +p^-_k(n,\hat c)),}
\eqno(2.4)
$$

\noindent where the sum is taken over all positive lattice directions,
i.e. $\hat a=1,2,3$. In the continuum limit, this goes over to

$$
D(n) \buildrel{a\rightarrow 0}\over\longrightarrow a^6 
\epsilon_{abc}\,\epsilon^{ijk} E^a_i E^b_j E^c_k+O(a^7).\eqno(2.5)
$$

To arrive at (2.4), we have substituted the continuum momenta $E^a_i$
by the averaged lattice momenta $p_i(n,\hat a):= 
\frac{1}{2} (p^+_i(n,\hat a) +p^-_i(n,\hat a))$, which ensures that
the final expression for $D(n)$ is invariant under relabelling of axes,
and no direction is preferred. This seems the only simple choice of
lattice function with these properties, and was the one adopted in [6]
to analyze some spectral properties of the volume operator. 

We may interpret ${\cal V}(n):=\sqrt{\frac{1}{6}D(n)}$ as the volume associated
with the  dual unit cube of $\Lambda^*$ centered at $n$, or as the lattice
version of $\det E$ at the point $n$ (there really is no distinction
between ``local" and ``smeared over a unit cube" on the lattice). The
latter interpretation is closer related to the continuum formulation, where
the finite operators (after the regulator has been removed) act
non-trivially at points $x\in\gamma$, or rather, on segments of the graph
$\gamma$ meeting at $x$. More generally, we define

$$
{\cal V}({\cal R})=\sum_{n\in{\cal R}} \sqrt{\frac{1}{3!}
D(n)} \eqno(2.6)
$$

\noindent for the volume of a lattice region $\cal R$ (a choice that 
will be justified in Sec.4).
There are several possibilities for discretizing local area.
Following the viewpoint of the dual lattice, the area of a unit two-surface
``perpendicular" to the $\hat a$-direction may be defined as

$$
{\cal A}_1(n,\hat a)=\sqrt{\frac{1}{2} 
(p^+_i(n,\hat a)p^{+i}(n,\hat a)+
p^-_i(n+1_{\hat a},\hat a)p^{-i}(n+1_{\hat a},\hat a))}\eqno(2.7)
$$

\noindent (no summation over $\hat a$), which should be interpreted as the
area of a unit surface  in $\Lambda^*$, dual to the link $l=(n,\hat a)$.
Alternatively, if one prefers to think of the operator action as  taking
place at the vertices $n$, one may define 
 
$$
{\cal A}_2(n,\hat a)=\sqrt{p_i(n,\hat a)p^i(n,\hat a)}\equiv
\sqrt{\frac{1}{4}(p^+_i(n,\hat a) 
+p^-_i(n,\hat a)) (p^{+i}(n,\hat a) +p^{-i}(n,\hat a))}\eqno(2.8)
$$
\noindent or
$$
{\cal A}_3(n,\hat a)=\sqrt{\frac{1}{2} 
(p^+_i(n,\hat a)p^{+i}(n,\hat a)+
p^-_i(n,\hat a)p^{-i}(n,\hat a))}.\eqno(2.9)
$$

The functional form of ${\cal A}_2$ is the one that appears as a special
case (i.e. when evaluated on states $\gamma$ that lie on an imbedded
lattice $\Lambda^{\rm imb}$ in $\Sigma$ of the area
operator in the continuum theory [10]). In the limit as $a\rightarrow 0$,
the expansions of the discretized area functions are all of the form 

$$
{\cal A}_i(n,\hat a)=a^2 \sqrt{E^{3i} E^3_i}\, +O(a^3).\eqno(2.10)
$$

\noindent For finite lattice areas $\cal S$, perpendicular to $\hat a$,
we define ${\cal A}_I({\cal S})=\sum_{n\in{\cal S}} {\cal A}_i (n,\hat a)$.

Due to its complicated functional form,
the definition of the length of a unit link is even more ambiguous. Thinking
in terms of link length  on the dual lattice, one would associate it with the
dual unit plaquette in $\Lambda$. To obtain a symmetric expression, one
possibility is to sum over the bi-vectors based at each of the four corners
of the plaquette. Thus the continuum expression
$E^a_i E^{ai} E^b_j E^{bj}-(E^a_i E^{bi})^2$ would (up to powers of $a$)
be represented by $\frac{1}{4}$ times the sum of the four terms

$$
\eqalign{
&C(n,\hat a,\hat b):=\cr
&\hskip.5cm p^+_i(n,\hat a)p^{+i}(n,\hat a)p^+_j(n,\hat b)p^{+j}(n,\hat b) -
(p^+_i(n,\hat a)p^{+i}(n,\hat b))^2,\cr
&C(n+1_{\hat a},-\hat a,\hat b):=\cr
&\hskip.5cm p^-_i(n+1_{\hat a},\hat a)p^{-i}(n+1_{\hat a},\hat a)
p^+_j(n+1_{\hat a},\hat b)p^{+j}(n+1_{\hat a},\hat b) -
(p^-_i(n+1_{\hat a},\hat a)p^{+i}(n+1_{\hat a},\hat b))^2,\cr
&C(n+1_{\hat a}+1_{\hat b},-\hat a,-\hat b):=\cr
&\hskip.5cm p^-_i(n+1_{\hat a}+1_{\hat b},\hat a)
p^{-i}(n+1_{\hat a}+1_{\hat b},\hat a)
p^-_j(n+1_{\hat a}+1_{\hat b},\hat b)p^{-j}(n+1_{\hat a}+1_{\hat b},\hat b) -
\cr
&\hskip2cm(p^-_i(n+1_{\hat a}+1_{\hat b},\hat a)
p^{-i}(n+1_{\hat a}+1_{\hat b},\hat b))^2,\cr
&C(n+1_{\hat b},\hat a,-\hat b):=\cr
&\hskip.5cm p^+_i(n+1_{\hat b},\hat a)p^{+i}(n+1_{\hat b},\hat a)
p^-_j(n+1_{\hat b},\hat b)p^{-j}(n+1_{\hat b},\hat b) -
(p^+_i(n+1_{\hat b},\hat a)p^{-i}(n+1_{\hat b},\hat b))^2,}\eqno(2.11)
$$

\noindent (no sums over $\hat a$, $\hat b$).
To obtain the complete expression for the link length, one still has
to divide by the density factor. Two different ways of symmetrizing lead
to

$$
\eqalign{
{\cal L}_1(n,\hat a,\hat b)&=\sqrt{\frac{3}{2} (C(n,\hat a,\hat b)D(n)^{-1}+
C(n+1_{\hat a},-\hat a,\hat b)D(n+1_{\hat a})^{-1}+}\cr
&\hskip1.2cm\overline{C(n+1_{\hat a}+1_{\hat b},-\hat a,-\hat b)
D(n+1_{\hat a}+1_{\hat b})^{-1}+
C(n+1_{\hat b},\hat a,-\hat b)D(n+1_{\hat b})^{-1}) },}\eqno(2.12)
$$
\noindent or
$$
\eqalign{
{\cal L}_2(n,\hat a,\hat b)&=\sqrt{
6\, (C(n,\hat a,\hat b)+C(n+1_{\hat a},-\hat a,\hat b)+
C(n+1_{\hat a}+1_{\hat b},-\hat a,-\hat b)+
C(n+1_{\hat b},\hat a,-\hat b))}\cr
&\hskip3cm\overline{(D(n)+D(n+1_{\hat a})+D(n+1_{\hat a}+1_{\hat b})
+D(n+1_{\hat b}))^{-1}}.}\eqno(2.13)
$$

\noindent Alternatively, a simpler version containing only link variables 
based at $n$ is given by 

$$
{\cal L}_3(n,\hat a,\hat b)=\sqrt{6\, D(n)^{-1}(p_i(n,\hat a)p^i(n,\hat a)
p_j(n,\hat b)p^j(n,\hat b)-(p_i(n,\hat a)p^i(n,\hat 
b))^2)},\eqno(2.14)
$$

\noindent where the averaged link momenta $p_i(n,\hat a)$ have been used.
In the limit as $a\rightarrow 0$, all three expressions have an
$a$-expansion of the form

$$
{\cal L}_i(n,\hat a,\hat b)=a\, \sqrt{\frac{1}{\det E} 
(E^a_j E^{aj} E^b_k E^{bk}- (E^{aj}E^b_j)^2\, ) }+O(a^2).\eqno(2.15)
$$

\noindent For finite curves $\cal C$ perpendicular to the $\hat a$- and
$\hat b$-directions, say, we use as an obvious definition
${\cal L}_i({\cal C})= \sum_{n\in{\cal C}} {\cal L}_i(n,\hat a,\hat b)$.
 
One may think of yet another way of discretizing length, namely, by 
expressing the term under the square root in terms of the dreibeins,
$g_{11}=e_{1}^{i}e_{1i}$, say, and using a lattice equivalent of the
continuum identity

$$
e_{a}^{i}\equiv\frac{1}{2\sqrt{\det E}}\,\epsilon_{abc}\epsilon^{ijk}
E_{j}^{b}E_{k}^{c}=2\,\big\{A_{a}^{i},\int d^{3}x\,\sqrt{\det E}\big\}.
\eqno(2.16)
$$

One motivation for this choice is the substitution of terms containing
negative powers of the square root of the metric by Poisson bracket
terms $\{A,\int\sqrt{\det E}\}$, which are not obviously singular
for vanishing $\det E$ [9]. As a discretized analogue one may use, for
example,

$$
-2\,\biggl\{ V(n,\hat a)_{A}{}^{B},\sum_{n}\sqrt{\frac{1}{6} D(n)}\biggr\} \,
V(n,\hat a)^{-1}{}_{B}{}^{C}\tau_{iC}{}^{A}\,
\buildrel{a\rightarrow 0}\over\longrightarrow 
\frac{a}{\sqrt{G}}\, e_{a}^{i}(x)+O(a^{2}).\eqno(2.17)
$$

\noindent The $g_{11}$-term can be discretized by the following 
expression:

$$
2\,\biggl\{V(n,\hat 1)_{A}{}^{B},\sum_{n}\sqrt{\frac{1}{6}D(n)}\biggr\}
\biggl\{V(n,\hat 
1)^{-1}{}_{B}{}^{A},\sum_{m}\sqrt{\frac{1}{6}D(m)}\biggr\}
\buildrel{a\rightarrow 0}\over\longrightarrow
\frac{a^{2}}{G}\, e_{1}^{i}e_{1i}\, +O(a^{3}). \eqno(2.18)
$$

However, it turns out that this construction is not particularly 
convenient on the lattice because of the appearance of the link 
holonomies $V(n,\hat a)$. Their quantum analogues, unlike the 
$\hat p$-operators, change the flux-line (or spin) assignments of the 
spin-network states they act upon. Equivalently, the quantized
expression (2.18) does not commute with the Laplacians (4.3), and 
therefore cannot be diagonalized on the same finite-dimensional 
sub-Hilbert spaces as the local geometric operators we have considered
so far. Unlike in the continuum, we cannot shrink away the link 
holonomies appearing in the quantum operators independently from those
appearing in the wave functions. Thus the analysis of this type of length 
operator is structurally more complicated, and we will not consider it 
presently. 

\vskip2cm

\line{\ch 3 Some explicit calculations\hfil}

In order to avoid unnecessary degeneracy of geometric operators, we 
will consider lattice spin-network states whose flux line assignments 
are non-vanishing. For our purposes it will be sufficient to study the 
local behaviour of such states around a vertex $n$. Following [5,6],
we will label the flux line assignments of the six links meeting at $n$
by a six-component vector $\vec j$, where $j_{1}$, $j_{2}$, $j_{3}$ 
correspond to the links in positive 1-, 2- and 3-direction, and 
$j_{4}$, $j_{5}$, $j_{6}$ to those in the three negative directions,
with $j_{i}=1,2,\dots$ (Fig.1). The value of $j_{i}$ is twice the
spin characterizing the irreducible representation of $SU(2)$ 
associated with the link.  
Recall that to specify a spin-network state 
locally, one needs in addition to $\vec j$ a choice of gauge-invariant
contractor for the flux lines meeting at $n$. Given $\vec j$, the 
space of all possible contractors at $n$ is finite-dimensional.

\hskip4.4cm\epsfxsize=150pt\epsfbox{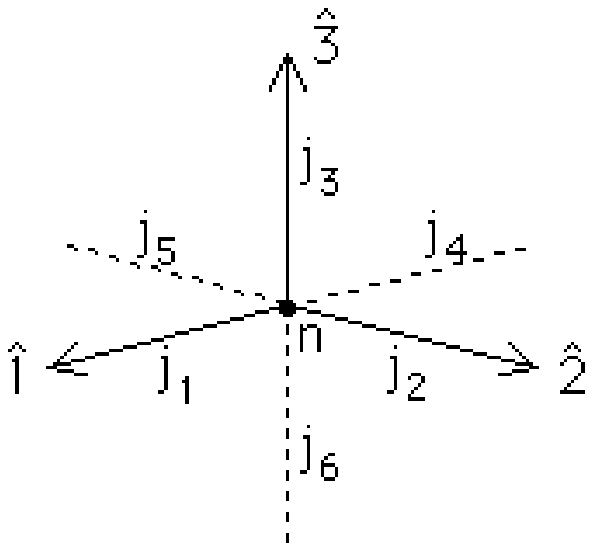}
\vskip0.1cm
\centerline{{\bf Fig.1}}

Let us consider the simplest type of a 6-valent intersection, 
namely, $\vec j=(1,1,1,1,1,1)$. 
There is one linear relation between the six spin-network states
$S_i$ (Fig.2) one can construct from these initial data, namely, 
$\tilde S:=S_1-S_2-S_3+S_4+S_5-S_6=0$. 
\vskip0.3cm
\epsfysize=90pt\epsfbox{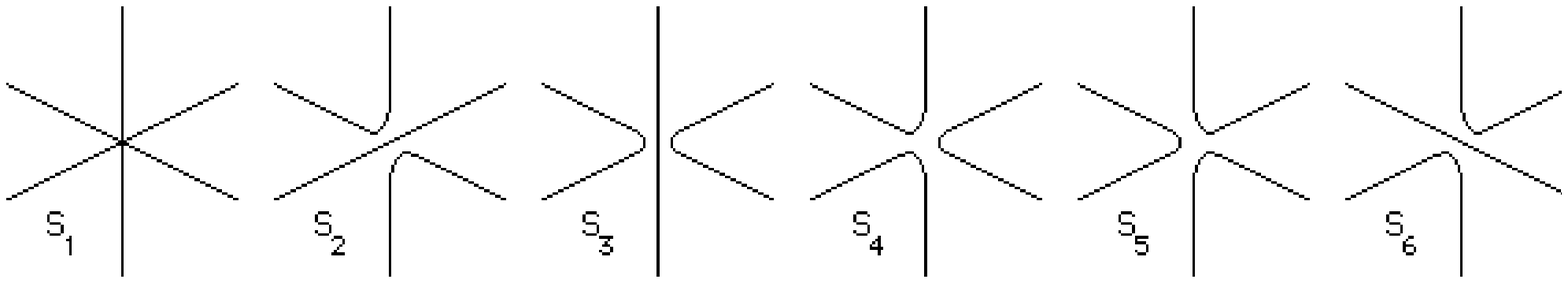}
\vskip0.1cm
\centerline{{\bf Fig.2}}
 
Using the area operators $\hat{\cal A}_2$ for measuring the local areas
in the three main directions, one finds a set of simultaneous eigenstates
with the following eigenvalues
\vskip1cm
{\offinterlineskip\tabskip=0pt
 \halign{ \strut\vrule#& \quad # \quad &\vrule#&
          \quad\hfil #\quad &\vrule#&
          \quad\hfil #\quad &\vrule#&
          \quad\hfil #\quad &\vrule#&
          \quad\hfil #\quad &\vrule#&
          \quad\hfil #\quad &\vrule#&
          \quad\hfil #\quad &\vrule#\cr \noalign{\hrule}
 & && && && && && && &\cr
 & && $S_1-2S_2$ && $S_1-2S_3$ && $S_1-2S_6$ 
              && $S_1$ && $S_4-S_5$ && $\tilde S$ &\cr
 & && && && && && && &\cr
 \noalign{\hrule}
 & && && && && && && &\cr
 & $\hat {\cal A}_2(n,\hat 1)$ &&
      $\sqrt{\frac{3}{4}}$ && $\sqrt{\frac{1}{4}}$ && $\sqrt{\frac{1}{4}}$ &&
      $\sqrt{\frac{3}{4}}$ && $\sqrt{\frac{1}{4}}$ && $\sqrt{\frac{1}{4}}$ &\cr
 & && && && && && && &\cr
 & $\hat {\cal A}_2(n,\hat 2)$ &&
      $\sqrt{\frac{1}{4}}$ && $\sqrt{\frac{1}{4}}$ && $\sqrt{\frac{3}{4}}$ &&
      $\sqrt{\frac{3}{4}}$ && $\sqrt{\frac{1}{4}}$ && $\sqrt{\frac{1}{4}}$ &\cr
 & && && && && && && &\cr
 & $\hat {\cal A}_2(n,\hat 3)$ && 
      $\sqrt{\frac{1}{4}}$ && $\sqrt{\frac{3}{4}}$ && $\sqrt{\frac{1}{4}}$ &&
      $\sqrt{\frac{3}{4}}$ && $\sqrt{\frac{1}{4}}$ && $\sqrt{\frac{1}{4}}$ &\cr
 & && && && && && && &\cr
 \noalign{\hrule} }
\normalbaselines
\baselineskip=16pt   
\vskip1cm
Using the area operators $\hat {\cal A}_3$, every state is an eigenstate,
with eigenvalue $\sqrt{\frac{3}{4}}$. 
Computing now the action of the operator $\hat D(n)$ on these states,
one obtains the following eigenvectors:

$$
\eqalign{
&{\rm eigenvalue }\hskip1.5cm 0:\hskip1cm S_1-2S_2,\; S_1-2S_3,\; 
  S_1-2S_6,\; \tilde S\cr
&{\rm eigenvalue }\hskip.8cm\frac{3}{2}\sqrt{\frac{3}{2}}:\hskip1cm
  \frac{1}{2}S_1 +S_{2}+S_{3}+(-1+\frac{i}{2}\sqrt{6})S_4 
  +(-1-\frac{i}{2}\sqrt{6})S_5+ S_{6}\cr
&{\rm eigenvalue }\hskip.4cm -\frac{3}{2}\sqrt{\frac{3}{2}}:\hskip1cm
  \frac{1}{2}S_1 +S_{2}+S_{3}+(-1-\frac{i}{2}\sqrt{6})S_4 
  +(-1+\frac{i}{2}\sqrt{6})S_5+ S_{6}. }
$$

Even taking into account that $\tilde S\sim 0$, the two eigenvectors
with eigenvalues $\pm\frac{3}{2}\sqrt{\frac{3}{2}}$ are not eigenvectors
of the areas $\hat {\cal A}_2$, i.e. $\hat D(n)$ and $\hat {\cal A}_2$
must necessarily be non-commuting operators. For simplicity, instead of
$[\hat {\cal V}(n), \hat {\cal A}_2(n,\hat 3)]$, we just calculate the
commutator of the polynomial expressions appearing under the square
roots (already omitting terms proportional to the Laplacian), i.e. 

$$
\eqalign{
[\hat D(&n),\hat p_i^+(n,\hat 3)\hat p^{-i}(n,\hat 3)]=
12\, i\, \hat p_i(n,\hat 1)\hat p_j(n,\hat 2) \hat p^{+[j}(n,\hat 3)
\hat p^{-i]}(n,\hat 3) =\cr
&6\,\epsilon^{ijk} \hat p_i(\hat 1)\hat p_j(\hat 2)\hat p_k^+(\hat 3)+
12\, i\, \hat p_i(\hat 1)\hat p_j(\hat 2) \hat p^{+[j}(\hat 3)
(\,\hat p^{+i]}(\hat 1) +\hat p^{+i]}(\hat 2)-\hat p^{-i]}(\hat 1)-
\hat p^{-i]}(\hat 2)\,),}\eqno(3.1) 
$$

\noindent which is not the zero-operator. 
In the second line we have not written out the vertex $n$ 
explicitly and have used the quantized version of
the discrete form of Gauss' law,

$$
p^+_i(n,\hat 1)+ p^+_i(n,\hat 2) + p^+_i(n,\hat 3)-
p^-_i(n,\hat 1)- p^-_i(n,\hat 2) - p^-_i(n,\hat 3)=0,
\eqno(3.2)
$$

\noindent to eliminate one of the six momentum operators based at $n$.
One easily checks that in the continuum limit, 
taking into account the expansions

$$
p_i^\pm (n,\hat b)=a^2\,G^{-1}\, E^b_i \pm a^3\, G^{-1}\,\nabla_b E^b_i+
O(a^4) \eqno(3.3)
$$

\noindent (no sum over $b$), the flux conservation relation (3.2) to lowest
order in $a$ is proportional to the usual expression for the Gauss law.

Now, {\it if} one were to define the volume operator by
$\hat{\cal V}=\sum \sqrt{\frac{1}{3!}|\hat D(n)|}$ (which however, as
we will explain in due course, is not the natural thing to do), one would
obtain a degenerate eigenspace for the eigenvalue
$+\frac{3}{2}\sqrt{\frac{3}{2}}$. It is easy to show that in the
example above (modulo the
zero-norm state $\tilde S$) a basis for this eigenspace is given by
$\{S_1, S_4-S_5\}$. Therefore, on the particular set of states with
$\vec j=(1,1,1,1,1,1)$, the operator $|\hat D(n)|$ does commute with all the
areas. Vanishing commutation relation with $|\hat D(n)|$ is a weaker
condition than with $\hat D(n)$ alone. The spectrum of $|\hat D(n)|$
coincides with the square root of the spectrum of $\hat D^2$, and
we have

$$
[\hat D^2,\hat P]=\hat D [\hat D,\hat P] +[\hat D,\hat P]\hat 
D.\eqno(3.4)
$$

\noindent It follows that for an arbitrary operator $\hat P$,
$[\hat D,\hat P]=0\Rightarrow [|\hat D|,\hat P]=0$, but not the other way
round. Substituting in the squared area operator, one obtains

$$
[\hat D^2,\hat p_i^+(\hat 3)\hat p^{-i}(\hat 3) ]=
2\, [\hat D,\hat p_i^+(\hat 3)\hat p^{-i}(\hat 3) ]\hat D \,
+{\rm higher-order\; terms\; in\;}\hbar.\eqno(3.5)
$$

\noindent We already know that $[\hat D,\hat p_i^+(\hat 3)\hat p^{-i}(\hat 3) ]$
is a non-vanishing operator, but it could in principle 
vanish on all states that are annihilated by $\hat D$ (in a volume eigenbasis), 
leading to $[\hat D,\hat p_i^+(\hat 3)\hat p^{-i}(\hat 3) ]\hat D=0$. 
However, the following example demonstrates that this does not happen.

\vskip0.3cm
\epsfysize=180pt\epsfbox{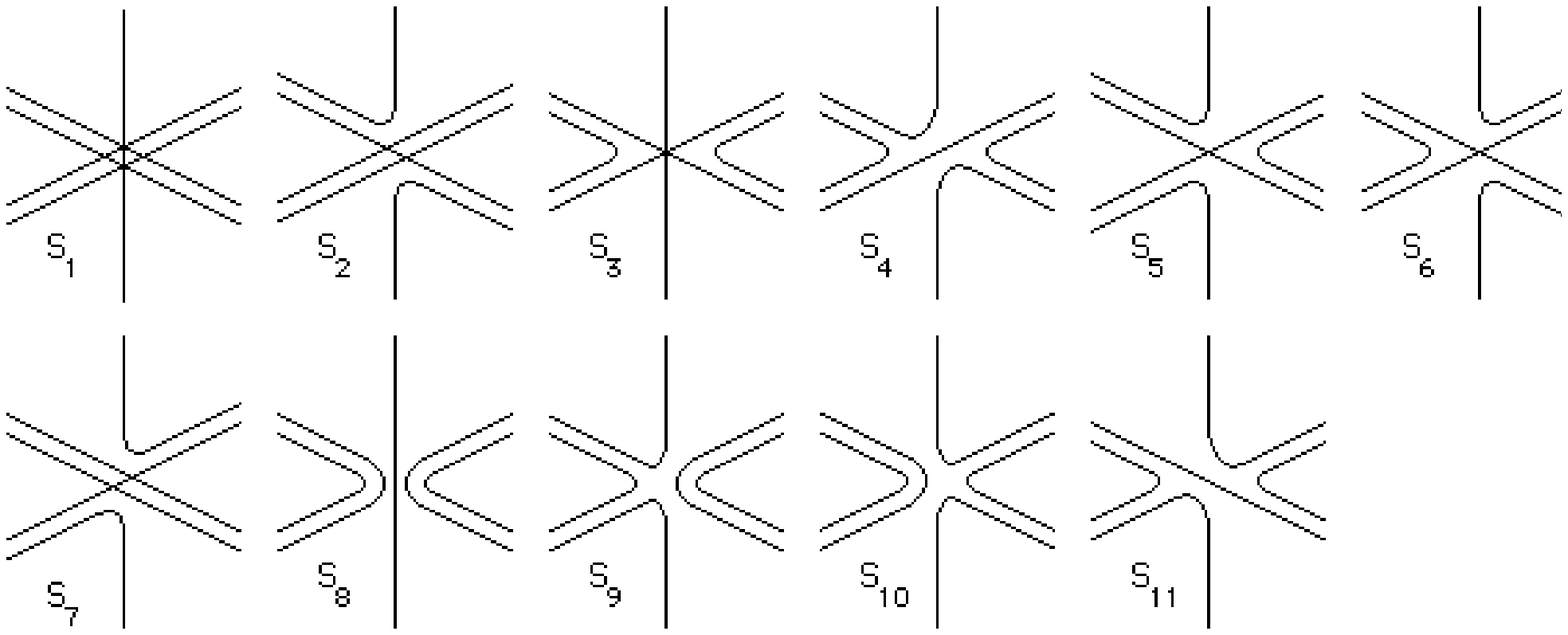}
\vskip0.1cm
\centerline{{\bf Fig.3}}

The simplest Hilbert subspace of loop configurations we have 
found where the commutator (3.4) is non-vanishing is one with 10 incoming 
flux lines at the vertex $n$, namely, $\vec j=(2,2,1,2,2,1)$. 
By taking appropriate linear combinations of
Wilson loop states (i.e. by antisymmetrizing over flux lines of 
multiply occupied links), one finds 11 spin-network states, subject to two 
Mandelstam identities, i.e. the subspace they span is 9-dimensional.
Let us call these states $\{ S_i,i=1,\dots ,11\}$, as illustrated in Fig.3. 
Simultaneous eigenstates of the three local area operators $\hat{\cal A}_2$ 
are the linear combinations $T_i$, defined by

$$
\eqalign{
&T_1=S_1-2S_7,\cr
&T_2=S_1-2S_2,\cr
&T_3=S_5-S_6,\cr
&T_4=S_1-2S_3,\cr
&T_5=S_1,\cr
&T_6=S_1-6S_3+6S_8,\cr
&T_7=S_1+S_2-3S_3-3S_5-3S_6-3S_7+12S_{11},\cr
&T_8=S_1-3S_2-3S_3+12S_4-3S_5-3S_6+S_7,\cr
&T_9=S_5-S_6-2S_9+2S_{10},\cr
&T_{10}=S_1-S_2-S_3+S_5+S_6-S_7,\cr
&T_{11}=S_3-S_4-S_8+S_9+S_{10}-S_{11}. }\eqno(3.6)
$$

\noindent Both $T_{10}$ and $T_{11}$ are zero-norm states, which will be
set identically to zero in what follows. The table of eigenvalues for the
remaining states is 
\vskip1cm
{\offinterlineskip\tabskip=0pt
 \halign{ \strut\vrule#& \quad # \quad &\vrule#&
          \quad\hfil #\quad &\vrule#&
          \quad\hfil #\quad &\vrule#&
          \quad\hfil #\quad &\vrule#&
          \quad\hfil #\quad &\vrule#&
          \quad\hfil #\quad &\vrule#&
          \quad\hfil #\quad &\vrule#&
          \quad\hfil #\quad &\vrule#&
          \quad\hfil #\quad &\vrule#&
          \quad\hfil #\quad &\vrule#\cr \noalign{\hrule}
 & && && && && && && && && && &\cr
 & && $T_1$ && $T_2$ && $T_3$ && $T_4$ && $T_5$ && 
      $T_6$ && $T_7$ && $T_8$ && $T_9$ &\cr
 & && && && && && && && && && &\cr
 \noalign{\hrule}
 & && && && && && && && && && &\cr
 & $\hat {\cal A}_2(n,\hat 1)$ && 
      $\sqrt{\frac{3}{2}}$ && $\sqrt{2}$ && $\sqrt{\frac{3}{2}}$ &&
      $\sqrt{\frac{3}{2}}$ && $\sqrt{2}$ && $\sqrt{\frac{1}{2}}$ &&
      $\sqrt{\frac{1}{2}}$ && $\sqrt{\frac{3}{2}}$ && $\sqrt{\frac{1}{2}}$ &\cr
 & && && && && && && && && && &\cr
 & $\hat {\cal A}_2(n,\hat 2)$ &&
      $\sqrt{2}$ && $\sqrt{\frac{3}{2}}$ && $\sqrt{\frac{3}{2}}$ &&
      $\sqrt{\frac{3}{2}}$ && $\sqrt{2}$ && $\sqrt{\frac{1}{2}}$ &&
      $\sqrt{\frac{3}{2}}$ && $\sqrt{\frac{1}{2}}$ && $\sqrt{\frac{1}{2}}$ &\cr
 & && && && && && && && && && &\cr
 & $\hat {\cal A}_2(n,\hat 3)$ &&
      $\sqrt{\frac{1}{4}}$ && $\sqrt{\frac{1}{4}}$ && $\sqrt{\frac{1}{4}}$ &&
      $\sqrt{\frac{3}{4}}$ && $\sqrt{\frac{3}{4}}$ && $\sqrt{\frac{3}{4}}$ &&
      $\sqrt{\frac{1}{4}}$ && $\sqrt{\frac{1}{4}}$ && $\sqrt{\frac{1}{4}}$ &\cr
 & && && && && && && && && && &\cr
 \noalign{\hrule} }
\normalbaselines
\baselineskip=16pt   
\vskip1cm
Calculating now the eigenvalues of the operator $\hat D(n)$ on the $T_i$,
one finds that the eigenvalue-zero space is five-fold degenerate.
The remaining eigenstates are

$$
\eqalign{
&{\rm eigenvalue }\hskip.8cm\frac{3}{2}\sqrt{\frac{5}{2}}:\hskip1cm
  V_1=5\, T_4 -i \sqrt{\frac{5}{2}}\, T_9\cr
&{\rm eigenvalue }\hskip.4cm -\frac{3}{2}\sqrt{\frac{5}{2}}:\hskip1cm
  V_2=5\, T_4 +i \sqrt{\frac{5}{2}}\, T_9\cr 
&{\rm eigenvalue }\hskip.8cm\frac{3}{2}\sqrt{\frac{23}{2}}:\hskip0.9cm
  V_3=8\, T_5 -3\, T_6 +3i \sqrt{\frac{23}{2}}\, T_3\cr
&{\rm eigenvalue }\hskip.4cm -\frac{3}{2}\sqrt{\frac{23}{2}}:\hskip0.8cm
  V_4=8\, T_5 -3\, T_6 -3i \sqrt{\frac{23}{2}}\, T_3.} 
$$

\noindent From the point of view of the operator $|\hat D(n)|$, the
sets $\{V_1,V_2\}$ and $\{V_3,V_4\}$ span the degenerate eigenspaces
with eigenvalues $\frac{3}{2}\sqrt{\frac{5}{2}}$ and
$\frac{3}{2}\sqrt{\frac{23}{2}}$ respectively. Still, the second of these
eigenspaces depends on more than two of the $T_i$, by which it is shown
that the $\hat {\cal A}_2$ and $|\hat D(n)|$ cannot be diagonalized 
simultaneously.

Let us now explain why {\it no} modulus should appear under the square
root in the definition of the volume operator. In order to
understand this, one has to go back to the definition of the
continuum canonical momentum $E_{i}^{a}(x)$\note{I thank T. Thiemann 
for a discussion on this point.}. This is a densitized
inverse dreibein, given by $E_{i}^{a}=\det e\, e^{a}_{i}$ where 
$e_{i}^{a}$ is the inverse dreibein satisfying

$$
e_{i}^{a}e_{aj}=\delta_{ij},\qquad e_{aj}e_{b}^{j}=g_{ab},\eqno(3.7)
$$

\noindent with the determinant $\det e$ taking values $\pm\sqrt{\det 
g}$. One therefore derives for classical, non-degenerate metrics
the inequality $\det E=(\det e)^4>0$. With $\det E$ positive, the
volume is simply $\int d^{3}x \,\sqrt{\det E}$. 

However, in Yang-Mills-type quantum representations with canonical
commutation relations

$$
[\hat A_{a}^{i}(x),\hat E_{j}^{b}(y)]=\delta_{a}^{b}\delta_{j}^{i}
\delta (x,y)\eqno(3.8)
$$

\noindent or commutation relations derived from (3.8), $\hat{\det E}>0$
is {\it not} automatic (indeed, in Yang-Mills phase space, there is
no such restriction). This is borne out by the fact that all non-zero
eigenvalues of $\hat D(n)$ seem to come in pairs $\pm d$ of opposite sign 
[6]. In gravity, we therefore have to {\it impose} a quantum analogue of 
$\det E>0$ in the quantum theory. On the lattice this is
straightforward -- go to a basis where all operators $\hat D(n)$ are
diagonal and eliminate all eigenstates with non-positive eigenvalues.
We do not know whether a similar restriction is compatible with the
continuum regularization of the volume operator. 

Let us briefly discuss the commutation relations of the simplest of
the length operators, $\hat {\cal L}_{3}$. We choose to abbreviate
its polynomial part by

$$
L(n,\hat a,\hat b):=
p_i(n,\hat a)p^i(n,\hat a)
p_j(n,\hat b)p^j(n,\hat b)-(p_i(n,\hat a)p^i(n,\hat b))^2\eqno(3.9)
$$

\noindent (no sums over $\hat a$ and $\hat b$), so that the entire
discretized length function is given by 
${\cal L}_{3}(n,\hat a,\hat b) =\sqrt{6\,D(n)^{-1}L(n,\hat a,\hat b)}$.
The commutator between $\hat D(n)$ and $\hat L(n,\hat a,\hat b)$ is
non-vanishing, but since its form is not particularly
illuminating, we will not write it explicitly. Note, however, that 
this implies a factor-ordering ambiguity in the definition of the
quantum operator $\hat {\cal L}_{3}(n,\hat a,\hat b)$, which was not 
present for the area and volume operators. The simplest explicit
case with $[\hat L(n,\hat a,\hat b),\hat D(n)]\not=0$ we have found
is the set of spin-network states with $\vec j=(2,1,1;2,1,1)$, and for
the set of states with $\vec j=(2,2,1;2,2,1)$ one finds that in addition also  
$[\hat L(n,\hat a,\hat b),\hat D(n)^{2}]\not=0$. 

In order not to have to address the factor-ordering problem, for the case of 
length and area functions, we confine ourselves to a computation of 
the classical Poisson brackets, which is identical with the quantum result to
lowest order in $\hbar$. Moreover, let us omit the square roots and
calculate first 

$$
\Big\{ \frac{ L(n,\hat 1,\hat 2) }{D(n)},p_{k}(\hat 3)p^{k}(\hat 3)
\Big\}=-\frac{ L(n,\hat 1,\hat 2) }{D(n)^{2}}\,
\{ D(n), p_{k}(\hat 3)p^{k}(\hat 3) \},\eqno(3.10)
$$

\noindent which by virtue of (3.5) is non-vanishing. Slightly more 
complicated is the computation of the Poisson brackets of the length 
in 3-direction and the area perpendicular to the
1-direction, say. It is proportional to 

$$
\eqalign{
&\Big\{ \frac{ L(n,\hat 1,\hat 2) }{D(n)},p_{k}(\hat 1)p^{k}(\hat 1)
\Big\}=-\frac{ L(n,\hat 1,\hat 2) }{D(n)^{2}}\,
\{ D(n), p_{k}(\hat 1)p^{k}(\hat 1) \}\,+\frac{1}{D(n)}
\{ L(n,\hat 1,\hat 2), p_{k}(\hat 1)p^{k}(\hat 1) \}\cr
&\hskip.1cm =-\frac{1}{D(n)^{2}}\bigl( 6\,L(n,\hat 1,\hat 2) p_{i}(
\hat 3)p_{j}(\hat 1) p^{[j+}(\hat 1) p^{i]-}(\hat 3) +
D(n)\, \epsilon_{jkl}  p_{(i}(\hat 2) p_{j)}(\hat 2)\,
p^{i+}(\hat 1) p^{l+}(\hat 1) p^{k-}(\hat 1) \bigr),}\eqno(3.11) 
$$

\noindent with the round brackets denoting symmetrization.
Assuming suitable regularity conditions (like, for instance, $D(n)>0$), 
we conclude that lengths and areas in general do not commute.


\vskip2cm

\line{\ch 4 Why does non-commutativity arise?\hfil}

Having seen some instances of non-commutativity of geometric operators
in the previous section, one may wonder whether this was to be expected.
Naively, it is rather surprising since the geometric operators contain
only information about components of the spatial metric, which classically
Poisson-commute. One obvious difference in the connection
approach is the fact that the classical dreibein variables $E^a_i$
(or composite quantities containing them) in the quantum theory are
represented by differential operators. This is the opposite of
what happens in quantum representations based on the metric formulation,
where the operators $\hat g_{ab}$ usually act by multiplication.

However, on the lattice, the non-commutativity we have found is not a
quantum effect, but simply a consequence of the regularization. Not
even the basic classical variables of lattice gauge theory obey the canonical 
Poisson brackets of their continuum counterparts. 
Only in the continuum limit as $a\rightarrow 0$ one rederives the
expected result. Consider, for example, the classical Poisson relation

$$
\{ p^+_i(n,\hat a), V_A{}^C(m,\hat b)\}=
-\frac{1}{2}\,\d_{nm}\d_{\hat a\hat b}\, \tau_{iA}{}^B V_B{}^C
(n,\hat a).\eqno(4.1)
$$

\noindent Replacing the link variables by their expansions (2.2), one obtains

$$
\{a^2 G^{-1} E^a_i+O(a^3),\frac{1}{2}\,a\,GA_b^j\tau_{jA}{}^C+O(a^2)\}=
-\frac{1}{2}\,\delta_{nm}\delta_{\hat a\hat b}\,\tau_{iA}{}^B (\one_B^C+O(a)).
\eqno(4.2)
$$

\noindent Dividing both sides by $a^3$ and using $\frac{1}{a^3}\delta_{mn}
\buildrel{a\rightarrow 0}\over\longrightarrow  
\delta^3 (x,y)$, one finds in the limit the canonical Poisson
brackets $\{E^a_i(x),A_b^j(y)\}=-\delta_{ij}\delta_b^a\delta^3(x,y)$.
Likewise, the non-vanishing lattice brackets $\{p^{\pm},p^{\pm}\}$ are
to lowest order in $a$ equivalent to the continuum brackets
$\{E,E\}=0$. In this case, the non-commutativity is clearly a
result of the regularization implicit in the definition of the basic 
lattice variables, and present both classically and 
quantum-mechanically. 

It is therefore not surprising when composite quantities depending on
link momenta do not commute in the lattice discretization. 
Sometimes one can find discretized versions of continuum expressions 
that have this property {\it before} the limit $a\rightarrow 0$ is taken.
It may then be convenient to use them, since in this case
a property of the continuum theory (commutativity of two functions) is 
implemented exactly on the lattice, which probably improves its
convergence behaviour. Typical examples are the sums of squares

$$
\sum_i
p^{\pm}_{i}(n,\hat a) p^{\pm\,i}(n,\hat a) \eqno(4.3)
$$ 

\noindent which commute with all other functions of the 
$p^{\pm}_{i}(m,\hat b)$'s. The corresponding quantum operators are of
course proportional to the Laplacian on the group manifold associated 
with the links $(n,\hat a)$ and $(n+1_{\hat a},\hat a)$ respectively. 

Applying the same reasoning to the Poisson bracket of the local volume and area 
function, one has schematically

$$
\{\sqrt{p^{3}},\sqrt{p^{2}}\}=\frac{1}{\sqrt{p^{5}}} \sum (\dots) p^{4}
\buildrel{a\rightarrow 0}\over\longrightarrow
\{\sqrt{E^{3}},\sqrt{E^{2}}\}=0.\eqno(4.4)
$$

\noindent To arrive at the second relation,
we have divided through by $a^{5}$ on both sides.
To summarize, the non-commutativity of discretized analogues of
geometric functions or their quantum operators is not a priori particularly
surprising or worrying. 

What conclusions may we draw from the above discussion for the 
continuum theory? 
It turns out that the calculations and results obtained in the lattice
discretization are special cases of calculations that may be done in the
continuum theory. That is, after choosing a particular geometric quantum 
operator $\hat O$ and a specific state $\psi$ in the continuum, the finite 
expression for $\hat O$ acting on $\psi$, after the regulator has been 
removed, may be identical with that of a particular lattice calculation
for a corresponding operator $\hat O^{\rm latt}$.

On the lattice, due to the discretization, one is more restricted in the 
kind of quantities one can compute at a ``point". For example, local area
can only be measured in the three directions given by the lattice axes
meeting at an intersection $n$. This happens because on the lattice there
is only a finite number of degrees of freedom associated with a unit cube,
namely, 9 (independent holonomy components) before and 6 after going to 
the quotient with respect to local gauge transformations. 
This corresponds to the 6 degrees of freedom
contained in the metric $g_{ab}$ before the imposition of diffeomorphism
symmetry. Hence there is a maximum of six independent metric quantities
that classically can be associated with a unit cell of the lattice. For
example, one may choose them to be the volume, three areas and two
length functions. 

Furthermore, due to the special geometry we have chosen for the lattice, 
we can only
compare graph or loop configurations that are at most six-valent (and
such that pairs of incoming links are collinear, and the entire set is not
coplanar at the intersection point). Obviously, in order to prove a statement 
like ``two operators do not commute", it is sufficient to exhibit a single 
instance of when they do not. For example, consider the area operator 
$\hat {\cal A}_2^{\rm cont}$ corresponding to some finite spatial two-surface 
$\cal S$ [10]. Now, take 
any continuum graph $\gamma$ with a six-valent intersection
(of the type just described) at some point $x\in\cal S$ such that  
two of the collinear pairs of incoming edges are tangent to $\cal S$ in
$x$, and such that the third pair in $x$ is collinear with the unit normal
defining the surface. For simplicity, the graph $\gamma$ is supposed to 
have no further intersections with itself or with $\cal S$ (in the present 
discussion, we ignore cases where global constraints prevent such a 
construction).

Evaluating the area operator on the finite set of spin-network states
associated with the graph $\gamma$ and some definite flux assignment 
to its edges, 
the only non-vanishing contributions come from the intersection at $x$,
and the calculation of eigenstates and eigenvalues is isomorphic to
the corresponding one on the lattice. On the same set of states one
may evaluate the volume operator of an arbitrary spatial region $\cal R$
containing $x$, as defined in [18], and again 
obtain the same result as one would have in the lattice calculation,
using $\hat {\cal V}$. It therefore follows immediately from our
calculations done in Sec.3 that $[\hat {\cal A}^{cont}({\cal S}), \hat 
{\cal V}^{cont}({\cal R})]\not= 0$ in the continuum. 

Considering next the local area operators associated with the three main 
directions on the lattice, one finds that they all mutually commute, independent 
of which of the three definitions we choose. We will not be bothered 
here with defining area functions for surfaces that lie obliquely in 
the lattice. There is no obstruction in principle to doing this, a
necessary requirement being that (in the limit as $a\rightarrow 0$) 
there are no preferred directions on the lattice. This is analogous 
with the restoration of rotational symmetry of observables in usual 
lattice gauge theory on a fixed Euclidean lattice in the continuum 
limit. 

In this restricted lattice framework we therefore 
cannot reproduce the result derived by Ashtekar, Corichi and Zapata, who
found non-commutativity when evaluating pairs of area operators on
certain configurations of spin-network states in the continuum [19]. 
The non-vanishing commutators between volumes and areas were found 
independently by the author during her lattice investigations. 
The origins of these different instances of non-commutativity are of 
course related.

As for the commutators involving the length functions, results of the 
lattice computations cannot be compared immediately to the only continuum 
result available [16], since the finite operators are not the same, as
explained earlier. However, we would find it rather surprising if the
continuum length operator defined there did commute with the volume, 
say. It is probably relatively straightforward to find instances of
explicit spin-network configurations (with non-vanishing volume) 
where they do not. After all, for an $n$-valent intersection, there 
can only exist a small number (of order $n$) of
independent, mutally commuting operators one may construct from the
right- and left-invariant vector fields on the corresponding $n$-fold
copy of $SU(2)$. 

One other comment concerns different definitions of the volume operator
in the continuum, which again have to do with the modulus sign under
the square root. In the version of the volume operator used in [7], in
evaluating the action on a spin-network state,
the modulus is taken for each individual term contributing to the sum 
at a given vertex, then the sum is taken, then the square root. By
contrast, the volume operator envisaged in
[18,8] has the modulus signs outside the entire sum, not each individual
term. (This and other differences between the two types of volume operators
are also discussed in [18].) The presence and different position of the 
modulus signs of course give rise to different finite operators, and in turn 
affect the 
quantum commutators with other geometric operators. Since {\it no}
modulus signs are present classically, it is a priori not even clear
that -- {\it if} one were to introduce something like a small-$a$
expansion also in the continuum -- the correct (vanishing) 
commutators could be reproduced. In this regard the lattice formulation
has a definite advantage, since one is not forced to introduce the
modulus anywhere, as we explained in the previous section.

The presence of anomalous commutators of geometric operators 
is a potentially worrying result from the point of view of the 
continuum theory, where there is no analogue of the lattice expansion 
parameter $a$, and it is usually claimed that no further 
continuum limits have to be taken [10,8] (for a criticism of this
approach, see the discussion 
in [20]). 
One could try to argue that the non-commutativity of classically
commuting quantities is a quantum effect, and therefore not unexpected.
Considering the lattice analogy, this is not really convincing:
the problem rather seems to be that the differential operators used to 
define the quantum geometric operators behave 
like non-local quantities, since their algebra with the non-local
holonomy variables is the same as that of the smeared-out non-local
lattice variables.

Note that a difference between the lattice and continuum approach is
that at a kinematical level (i.e. before spatial diffeomorphisms are
taken into account -- this is the usual setting for discussing 
geometric operators), in the lattice regularization there is no problem
in defining operators measuring local information about the metric. 
In the continuum, on the contrary, because of the way the quantum 
theory has been set up, only operators corresponding to {\it finite} areas 
and volumes can be regularized and therefore defined properly.

This does not invalidate our discussion about the commutativity or otherwise 
of geometric operators, since above we have only exploited that certain
calculations in both approaches are identical. 

One may also define length, area and volume operators for entire lattice regions,
in analogy with the continuum construction. 
For the case of the volume function this is straightforward [6]: 
simply sum over all vertices $n$ contained in the lattice region $\cal R$,

$$
{\cal V}({\cal R}_{\rm latt})=\sum_{n\in {\cal R}_{\rm latt}} 
\sqrt{ \frac{1}{3!} D(n) }.\eqno(4.5)
$$

\noindent When expanding in powers of $a$, this becomes

$$
\sum_{{\cal R}_{\rm latt}} a^3\, (\sqrt{\frac{1}{3!}
\epsilon_{abc}\,\epsilon^{ijk} E^a_i E^b_j E^c_k} +O(a)\, )
\buildrel{a\rightarrow 0}\over\longrightarrow
\int_{\cal R}d^3x\,\sqrt{\frac{1}{3!}
\epsilon_{abc}\,\epsilon^{ijk} E^a_i E^b_j E^c_k}.\eqno(4.6)
$$

\noindent For the area function, things are simplest when the area
${\cal S}_{\rm latt}$ coincides with some planar surface of lattice 
plaquettes. Then one has (for an area dual to the 3-direction)

$$
{\cal A}_{i}({\cal S}_{\rm latt})=\sum_{n\in {\cal S}_{\rm latt}}
{\cal A}_{i}(n,\hat 3)
\buildrel{a\rightarrow 0}\over\longrightarrow 
\sum_{{\cal S}_{\rm latt}} a^2\, (\sqrt{E^3_i E^{3i}}+O(a)\,).\eqno(4.7)
$$

\noindent Likewise, to obtain the length of a finite lattice curve ${\cal 
C}_{\rm latt}$, one possibility is to add up the contributions from the 
initial points of the individual links of ${\cal C}_{\rm latt}$. For 
the simple case of a curve along the 1-direction, this results in

$$
{\cal L}_{i}({\cal C}_{\rm latt})=\sum_{(n,\hat 1)} 
{\cal L}_{i}(n,\hat 2,\hat 3)\buildrel{a\rightarrow 0}\over\longrightarrow 
\sum_{(n,\hat 1)} a\,\Big(
\sqrt{\frac{1}{\det E} 
(E^2_j E^{2j} E^3_k E^{3k}- (E^{2j}E^3_j)^2\, ) 
}+O(a)\,\Big).\eqno(4.8)
$$

\noindent One observes that in all three cases the lowest power of $a$
in the expansion is just the correct one to saturate the relevant
continuum integral. 
Obviously these definitions are not complete without 
specifying how points along boundaries of spatial regions are to be
counted, but for our present discussion we are not interested in
spelling out the details of how this may be done.

Let us now demonstrate that on the lattice also the commutator of
a {\it finite} area with a {\it finite} volume vanishes in the continuum limit. 
Consider, for example,
the Poisson brackets between (4.5) and (4.7). Abbreviating
$C(n,\hat 3):= p_i(n,\hat 3) p^i(n,\hat 3)$, one has

$$
\eqalign{
\Bigl\{ \sum_{n\in {\cal S}_{\rm latt}}&\sqrt{ C(n,\hat 3)},
\sum_{n\in {\cal R}_{\rm latt}} \sqrt{ \frac{1}{3!} D(n) } \Bigr\} = 
\,\sum_{n\in {\cal S}_{\rm latt}\cap {\cal R}_{\rm latt}}
C(n,\hat 3)^{-\frac{1}{2}} D(n)^{-\frac{1}{2}}
\{ C(n,\hat 3),D(n) \} \cr
&\buildrel{a\rightarrow 0}\over\longrightarrow
\sum_{n\in {\cal S}_{\rm latt}\cap {\cal R}_{\rm latt}}
a^3\, ((E^5)^{-\frac{1}{2}}E^4 +O(a)\,). 
}
\eqno(4.9)
$$

\noindent Since the intersection ${\cal S}_{\rm latt}\cap 
{\cal R}_{\rm latt}$ describes (if it is non-vanishing) a two-dimensional
surface in the continuum, we can substitute 
$\sum_n a^2 \rightarrow \int d^2x$, and the right-hand side of (4.9)
becomes a two-dimensional integral over a function that vanishes as
$a\rightarrow 0$, i.e. goes itself to zero in this limit. Hence 
the fact that we computed commutators of the local instead of
the integrated geometric functions did not alter our main argument.

\vskip2cm
\line{\ch 5 Summary and discussion\hfil}

The conclusion of our computations in the previous sections is that
geometric operators, as they are usually defined in the
continuum loop representation for quantum gravity, in general do not commute, 
although they do classically -- being functions of half of the canonical 
variables only. We have shown this explicitly for the case of volume and 
area operator because in this case the lattice calculations could be directly
related to constructions already available in the continuum theory. 
However, our analysis suggests that this is a general feature of 
operators constructed along similar lines in the continuum 
representation, and may therefore also spoil other commutation
relations, such as those involving a suitably regularized length operator 
or Hamiltonian operator. 

We have argued that the non-preservation of classical commutators
is not primarily a quantum effect, since the same algebraic 
relations can already be obtained in a classical, discretized version 
of the theory, where all basic variables are constrained to live on
one-dimensional lattice links. 

On the other hand, non-commutativity in the lattice-regularized theory is 
not surprising, since the basic variables are by construction non-local,
and their Poisson algebra is not canonical. The usual canonical 
commutation relations are only obtained in the limit as the lattice
spacing $a$ is taken to zero. As we have shown, in this limit all
the geometric functions (volumes, areas, lengths) become Poisson-commuting,
as one would expect. At least for the case of volume and area, the same is 
also true for the corresponding quantum operators. 

Is there anything one could do in the continuum loop representation
to achieve commutativity? 
It does not seem that quotienting out by spatial diffeomorphisms would 
remedy the situation, since the action of the geometric operators is 
encoded in topological reroutings at graph intersections, which behave 
covariantly under spatial diffeomorphisms.
One may try to select a subspace of the Hilbert space spanned by the 
spin-network states on which the geometric operators commute, in which
case one would have to show that it is non-trivial (for example, that 
it contained states with non-vanishing volume).

Alternatively, one may interpret the result as an indication that it is
after all necessary to take some sort of continuum limit, even if the
ingredients of the quantum theory have been defined using a continuous
``background'' differential manifold. In keeping with the general spirit 
of this
ansatz, one would perhaps need some small ``topological" expansion parameter,
presumably coming from some topological characteristics of the quantum
states, for example, counting intersections of a certain type. 
(Further support for the need of a continuum limit comes from the 
following consideration.
The construction of geometric (and other) operators in the continuum
loop quantization is not universal. Since a small-distance 
expansion is always invoked to motivate the functional form of a
regularized operator, in principle the same type of ambiguity as in the lattice 
theory appears. In the latter case, since the ambiguity is the result 
of a choice of the discretization (i.e. regularization), 
a necessary condition one has to impose on the continuum limit is 
that different discretizations 
must lead to equivalent quantum theories. A similar criterion seems to be
missing so far in the continuum construction.) 

The lattice-regularized theory does not share this problem,
although one may of course argue that it has not yet been shown that
it leads to a sensible continuum quantum theory.  
The lattice approach has difficulties of its own, as there is no obvious way 
of implementing spatial diffeomorphism symmetry at the discretized level.
This can for example be done ``to lowest order in $a$", in which case one 
must show that the algebra of the diffeomorphism constraint does not acquire 
anomalous terms in the quantization. We are now in a state to tackle this
rather involved computation and hope to be able to report soon on its 
outcome [21].

\ni{\it Acknowledgement.} I am grateful to the organizers of the
Vienna workshop on quantum gravity for inviting me, and to its participants 
for useful comments and discussions.

\vskip2cm

\line{\ch References\hfil}

\item{[1]} C. Rovelli and L. Smolin: Loop space representation of
  quantum general relativity, {\it Nucl. Phys.} B331 (1990) 80-152.

\item{[2]} J.F. Barbero G.: Real Ashtekar variables for Lorentzian
  signature space-times, {\it Phys. Rev.} D51 (1995) 5507-10.

\item{[3]} A. Ashtekar: New variables for classical and quantum
  gravity, {\it Phys. Rev. Lett.} 57 (1986) 2244-7; A new 
  Hamiltonian formulation of general relativity, {\it Phys.
  Rev.} D36 (1987) 1587-1603.
  
\item{[4]} C. Rovelli and L. Smolin: Discreteness of area and 
  volume in quantum gravity, {\it Nucl. Phys.} B442 (1995)
  593-622, Err. {\it ibid.} B456 (1995) 753-4.

\item{[5]} R. Loll: The volume operator in discretized quantum
  gravity, {\it Phys. Rev. Lett.} 75 (1995) 3048-51.
  
\item{[6]} R. Loll: Spectrum of the volume operator in
  quantum gravity, {\it Nucl. Phys.} B460 (1996) 143-54.

\item{[7]} R. De Pietri and C. Rovelli: Geometry eigenvalues
  and scalar product from recoupling theory in loop quantum gravity,
  {\it Phys. Rev.} D54 (1996) 2664-90.

\item{[8]} R. Loll: A real alternative to quantum gravity in loop
  space, {\it Phys. Rev.} D54 (1996) 5381-4.

\item{[9]} T. Thiemann: Anomaly-free formulation of nonperturbative 
  four-dimensional Lorentzian quantum gravity, {\it Phys. Lett.} B380
  (1996) 257-64; Quantum spin dynamics I\&II, Harvard U. 
  {\it preprints} HUTMP-96/B-351 and B-352.

\item{[10]} A. Ashtekar and J. Lewandowski: Quantum theory of 
  geometry I: area operators, Penn State U. {\it preprint} 
  CGPG-96-2-4.
  
\item{[11]} S. Frittelli, L. Lehner and C. Rovelli: The complete 
  spectrum of the area from recoupling theory in loop quantum gravity,
  {\it Class. Quant. Grav.} 13 (1996) 2921-32.  

\item{[12]} see, for example, M. Barreira, M. Carfora and C. Rovelli:
  Physics with nonperturbative quantum gravity: radiation from a
  quantum black hole, Pittsburgh U. {\it preprint}, 1996.  

\item{[13]} see, for example, A. Ashtekar, C. Rovelli and L. Smolin:
  Weaving a classical geometry with quantum threads,
  {\it Phys. Rev. Lett.} 69 (1992) 237-40; L. Smolin:
  Recent developments in nonperturbative quantum gravity, in {\it Quantum
  Gravity and Cosmology}, ed. J. P\'erez-Mercader et al, World
  Scientific, Singapore, 1992, 3-84.

\item{[14]} C. Rovelli and L. Smolin: Spin networks and quantum 
  gravity, {\it Phys. Rev.} D52 (1995) 5743-59;
  J.B. Baez: Spin network states in gauge theory, to
  appear in {\it Adv. Math.};
  Spin networks and nonperturbative quantum gravity, 
  UC Riverside {\it preprint}, 1995.

\item{[15]} P. Renteln and L. Smolin: A lattice approach to spinorial
  quantum gravity, {\it Class. Quant. Grav.} 6 (1989) 275-94;
  P. Renteln: Some results of SU(2) spinorial lattice
  gravity, {\it Class. Quant. Grav.} 7 (1990) 493-502;
  O. Bostr\"om, M. Miller and L. Smolin: A new discretization of
  classical and quantum general relativity, Syracuse U. {\it preprint} 
  SU-GP-93-4-1;
  R. Loll: Non-perturbative solutions for lattice quantum gravity,
  {\it Nucl. Phys.} B444 (1995) 619-39;
  K. Ezawa: Multi-plaquette solutions for discretized 
  Ashtekar gravity, {\it Mod. Phys. Lett.} A11 (1996) 2921-32;
  H. Fort, R. Gambini and J. Pullin: Lattice knot theory and quantum
  gravity in the loop representation, Penn State U. {\it preprint}
  CGPG-96/8-1.

\item{[16]} T. Thiemann: A length operator for canonical quantum 
  gravity, Harvard U. {\it preprint} HUTMP-96/B-354.

\item{[17]} J. Kogut and L. Susskind: Hamiltonian formulation of
  Wilson's lattice gauge theories, {\it Phys. Rev.} D11 (1975)
  395-408; J.B. Kogut: The lattice gauge theory approach
  to quantum chromodynamics, {\it Rev. Mod. Phys.} 55 (1983) 775-836.

\item{[18]} J. Lewandowski: Volume and quantizations, 
  Warsaw U. {\it preprint}, 1996. 

\item{[19]} A. Ashtekar: talk given at ESI workshop, July 1996.

\item{[20]} L. Smolin: The classical limit and the form of the 
  hamiltonian constraint in nonperturbative quantum gravity,
  Penn State U. {\it preprint} CGPG-96/9-4.

\item{[21]} R. Loll, in preparation.

\end